\documentclass{aa}         
\usepackage{graphicx}
\usepackage{epsfig}
\usepackage{natbib}
\bibpunct{(}{)}{;}{a}{}{,}
\begin{document}    
\title{The ELODIE survey for northern extra--solar planets\thanks{Based on observations 
made at the Haute--Provence Observatory (operated by French CNRS), the 1.2--m Euler 
swiss telescope at {\footnotesize ESO}--La Silla Observatory (Chile) and  the 1.52--m
{\footnotesize ESO}\ telescope also at La Silla Observatory.} \thanks{The 
{\footnotesize ELODIE} measurements discussed in this paper are only available in 
electronic form at the {\footnotesize CDS} via anonymous ftp to {\tt 
cdsarc.u-strasbg.fr (130.79.125.5)} or via 
{\tt http://cdsweb.u-strasbg.fr/Abstract.html}}}

\subtitle{I. 6 new extra--solar planet candidates}

\author{C.~Perrier\inst{1} \and  J.-P.~Sivan\inst{2} \and D.~Naef\inst{3} \and 
J.L.~Beuzit\inst{1} \and M.~Mayor\inst{3} \and D.~Queloz\inst{3} \and S.~Udry\inst{3}}

\institute{Laboratoire d'Astrophysique de Grenoble, Universit\'e J. Fourier, BP 53, 
F--38041 Grenoble, France \and Observatoire de Haute-Provence, F--04870 St-Michel 
L'Observatoire, France \and Observatoire de Gen\`eve, 51 ch. des Maillettes, CH--1290
Sauverny, Switzerland}

\offprints{Christian Perrier, \email{christian.perrier@obs.ujf-grenoble.fr}}

\date{Received / Accepted}

\abstract{
Precise radial--velocity observations at Haute--Provence Observatory 
({\footnotesize OHP}, France) with the {\footnotesize ELODIE} echelle spectrograph 
have been undertaken since 1994. In addition to several discoveries described 
elsewhere, including and following that of 51\,Peg\,b, they reveal new sub--stellar 
companions with essentially moderate to long periods. We report here about such 
companions orbiting five solar--type stars (\object{{\footnotesize HD}\,8574}, 
\object{{\footnotesize HD}\,23596}, \object{{\footnotesize HD}\,33636}, \object{{\footnotesize HD}\,50554}, 
\object{{\footnotesize HD}\,106252}) and one sub--giant star (\object{{\footnotesize HD}\,190228}). 
The companion of \object{{\footnotesize HD}\,8574} has an intermediate period of 227.55\,days 
and a semi--major axis of 0.77\,AU. 
All other companions have long periods,  exceeding 3\,years, and consequently  
their semi--major axes are around or above 2\,AU. The detected companions have 
minimum masses $m_{\rm 2} \sin i$ ranging from slightly more than 2\,M$_{\rm Jup}$
to  10.6\,M$_{\rm Jup}$. These additional objects reinforce the conclusion that most 
planetary companions have masses lower than 5\,M$_{\rm Jup}$ but with a tail of the 
mass distribution going up above 15\,M$_{\rm Jup}$. 
The orbits are all eccentric and 4 out of 6 have an  eccentricity of the order 
of 0.5. 
Four stars exhibit solar metallicity, one is metal--rich and one metal--poor. 
With 6 new extra--solar planet candidates discovered, increasing their total 
known to--date number to 115, the {\it {\footnotesize ELODIE} Planet Search 
Survey} yield is currently 18. 
We emphasize that 3 out of the 6 companions could in principle be resolved 
by diffraction--limited imaging on 8\,m--class telescopes depending on the achievable 
contrast, and therefore be primary targets for first attempts of extra--solar planet 
direct imaging.
}

\titlerunning{The {\footnotesize ELODIE} survey for northern extra-solar planets I}
\authorrunning{C.~Perrier et al.} 
\maketitle

\section{Introduction}

The {\it {\footnotesize ELODIE} Planet Search Survey}, an extensive radial--velocity 
northern survey of dwarf stars, has been underway for several years at the 
Haute--Provence Observatory ({\footnotesize OHP}, {\footnotesize CNRS}, France) 
using the {\footnotesize ELODIE} high-precision fiber-fed echelle spectrograph 
\citep{Baranne96} mounted on the Cassegrain focus of the 1.93--m telescope. 
This survey was initiated in 1994 by M. Mayor and D. Queloz in order to detect very 
low-mass stellar companions and allowed them to discover the first extra--solar planet 
(\object{51\,Peg\,b}) orbiting a solar-type star \citep{MayorQueloz95}. 
This discovery was followed by a number of subsequent detections of extra--solar planets 
with {\footnotesize ELODIE}. Some of these results have previously been published.

This survey is part of a large effort aiming at extra--solar planet search through 
radial--velocity measurements since several other planet surveys with a sensitivity 
$K_{\rm 1}$\,$>$\,10\,m\,s$^{\rm -1}$\ are underway elsewhere. 
A list of the most productive ones includes the {\em Geneva Southern Planet 
Search Programme}, carried out with the {\footnotesize CORALIE} spectrograph mounted 
on the 1.2--m Euler swiss telescope \citep{QuelozMW00,Udry00} at ESO--La Silla 
observatory, the {\em California \& Carnegie Planet Search} with the {\footnotesize 
HAMILTON} spectrograph at the Lick Observatory \citep{Marcy92} and the {\footnotesize 
HIRES} spectrograph mounted on the 10--m Keck--1 telescope \citep{Vogt94} at the 
W.~M.~Keck Observatory, the {\em G-Dwarf Planet Search} also performed with the 
{\footnotesize HIRES} spectrograph at Keck Observatory \citep{Latham00}, the 
{\em Anglo-Australian Planet Search} with the {\footnotesize UCLES} echelle 
spectrograph mounted on the 3.92--m Anglo-Australian Telescope \citep{Tinney01}, the 
{\sl {\footnotesize AFOE} Planet Search} \citep{KorzennikCSSSS} using the 
{\footnotesize AFOE} spectrograph mounted on the 1.5--m telescope at the Whipple Observatory and 
the {\sl McDonald Planetary Search} \citep{Cochran94} using the McDonald Observatory 2.7--m telescope coude spectrograph.

115 planet candidates, including 11 planetary systems, with  minimum masses 
lower than 18\,M$_{\rm Jup}$ and down to sub-Saturnian planets \citep{Marcy00,Pepe02}, 
have been detected so far, with periods from slightly less than 3 days to about 15 
years. 
This should permit us to obtain more robust conclusions about the mass function 
estimation, and notably its fast rise towards lower masses, the relative metal content 
of the parent stars and constraints on the so-called {\it brown dwarf desert} at 
separations of a few AU \citep{Halbwachs00} and on the emerging {\it period desert} 
between 10 and 70\,days \citep{Udrywash}, crucial ingredients to the understanding of 
planet formation mechanisms.

\begin{table}[t!]       
\caption{The {\bf 18} planet candidates with $m_{\rm 2}\sin i$ $<11$ M$_{\rm 
Jup}$\ detected with {\footnotesize ELODIE} at Haute--Provence Observatory
}       
\begin{tabular}{ll}
\hline\hline
\noalign{\vspace{0.025cm}}
{\it Candidate}                                      & {\it References} \\
\noalign{\vspace{0.025cm}}
\hline
\object{51\,Peg\,b}                                           & \citet{MayorQueloz95}\\
\object{14\,Her\,b}                                           & \citet{UdryMQ00}, \citet{NaefeloIII}\\
\object{{\footnotesize Gl}\,876\,b}$^{\bullet}$               & \citet{Delfosse98}\\
\object{{\footnotesize HD}\,209458\,b}$^{\diamond}$           & \citet{Mazeh00}\\
\object{{\footnotesize HD}\,190228\,b}$^{\star}$              & \citet{Sivan00} and this paper\\
\object{{\footnotesize HD}\,8574\,b}$^{\star}$                & this paper\\
\object{{\footnotesize HD}\,50554\,b}$^{\star}$$^{\dag}$      & this paper\\
\object{{\footnotesize HD}\,74156\,b}$^{\star}$               & \citet{NaefeloIII}\\
\object{{\footnotesize HD}\,74156\,c}$^{\star}$               & \citet{NaefeloIII}\\
\object{{\footnotesize HD}\,80606\,b}$^{\star}$$^{\circ}$     & \citet{Naef01}\\
\object{{\footnotesize HD}\,106252\,b}$^{\star}$$^{\dag}$     & this paper\\
\object{{\footnotesize HD}\,178911\,Bb}$^{\circ}$$^{\star}$   & \citet{Zucker02}\\
\object{{\footnotesize HD}\,20367\,b}$^{\ast}$                & \citet{Udrywash}\\
\object{{\footnotesize HD}\,23596\,b}$^{\ast}$                & this paper\\
\object{{\footnotesize HD}\,33636\,b}$^{\ast}$$^{\ddag}$      & this paper\\
\object{{\footnotesize HD}\,37124\,c}$^{\ast}$$^{\natural}$   & \citet{Udrywash}\\
\object{{\footnotesize HD}\,150706\,b}$^{\ast}$               & \citet{Udrywash}\\
\object{{\footnotesize GJ}\,777\,Ab}$^{\ast}$$^{\flat}$       & \citet{Udrywash, NaefGJ777}\\
\noalign{\vspace{0.025cm}} 
\hline       
\end{tabular}       
\begin{list}{}{}
\item$^{\bullet}$\ See also \citet{Marcy98}
\item$^{\diamond}$\ See also \citet{Charbonneau00,Henry00}	    
\item$^{\star}$\ Announced in {\footnotesize ESO} Press Release nb 07/01 (April 4, 2001)
\item$^{\dag}$\ Confirmed by \citet{Fischer02}             
\item$^{\circ}$\ Variability detected by the Keck {\em G--Dwarf Planet Search}	    
\item$^{\ddag}$\ Independently announced by \citet{Vogt02}      
\item$^{\ast}$\ Announced during the {\sl Scientific Frontiers in Research on 
extra--solar Planets} conference, Washington, June 2002
\item$^{\natural}$\ See also \citet{Butler02}
\item$^{\flat}$\ {\footnotesize HD}\,190360\,b
\end{list}    
\label{decouvertesElodie}
\end{table}

This paper is the first of a series devoted to the discovery of extra--solar planet 
candidates in the framework of the {\it {\footnotesize ELODIE} Planet Search Survey}. 
It reports on the detection of six new candidates, with minimum planetary masses 
($m_{\rm 2} \sin i$) in  the range 2 to 11 Jovian masses. 
Five of these are orbiting F8 or G0 dwarf stars and one is found around a G5IV 
star. 
They are all but one long-period objects on little to moderately elongated orbits: 
three periods are found to be greater than 1500 days, the shortest one is 227.55 days 
and the orbital eccentricities are of the order of 0.5 for four objects and
close to 0.3 for the two other ones. 
These discoveries, together with 3 others announced  during the {\sl Scientific 
Frontiers in Research on extra--solar Planets} conference in Washington (June 2002), 
bring the number of planets discovered with {\footnotesize ELODIE} up to 18.

The next section describes the sample of the {\it {\footnotesize ELODIE} Planet 
Search Survey} and summarizes the planet candidates it has allowed to detect up 
to now. 

The basic properties of the stars harboring the six new companions are given in 
Sect.~\ref{secStellarProperties}. 
Sect.~\ref{secSolOrbitales} presents the radial--velocity data and the orbital 
solutions for the companions. The case of \object{{\footnotesize HD}\,13507}, 
suspected for some time to harbor a planet but that actually has a low mass star 
companion, is exposed in Appendix~A.


\section{The {\footnotesize ELODIE}\ Planet Search Survey}\label{secElodiePSS}

\begin{figure}[t!]    
\psfig{figure=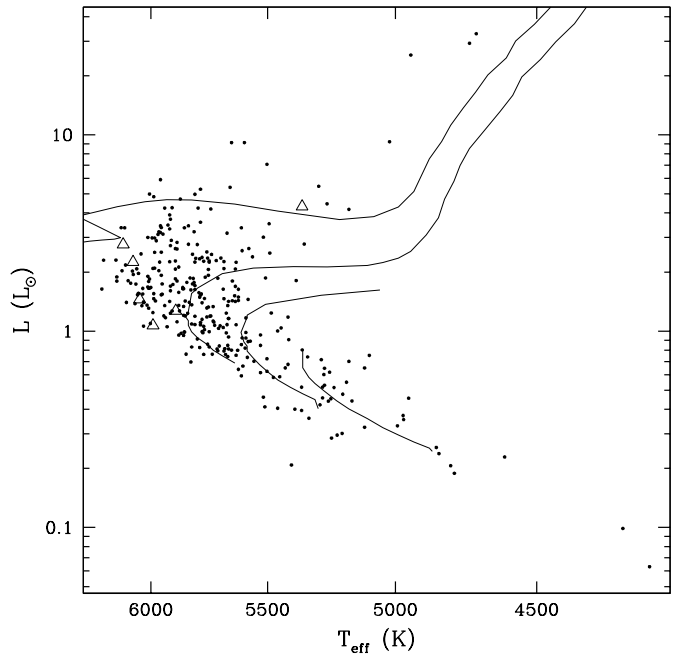,width=1\hsize}
\caption{
\label{diagramHR}    
The Hertzsprung-Russel diagram for the {\footnotesize ELODIE} main programme sample 
stars. The six stars with new planet candidates are shown as open triangles. 
The four curves are the solar metallicity evolutionary tracks for 0.8, 0.9, 1.0 and 
1.25\,M$_{\sun}$ from Geneva models \citep{Schaller92} 
}
\end{figure}

The original sample of \citet{MayorQueloz95} contained 142 stars, out of which \object{51\,Peg}. 
The main selection criterion of these G and K dwarfs was their radial--velocity 
non--variability according to the anterior {\footnotesize CORAVEL} survey 
\citep{DM91I,DM91II}.
The sample was largely modified in 1996 and later, so that the to--date survey sample 
size amounts up to 372 stars. 
Among the present sample, only 71 stars remain from the original list (active and/or 
too faint stars of this list were excluded) and 259 additional stars were included 
with the following criteria: $m_{\rm V}$\,$\leq$\,7.65, $\delta$\,$>$\,0\,$\degr$, 
$v \sin i$ (from {\footnotesize CORAVEL}) $<$\,4 km\,s$^{\rm -1}$, {\footnotesize 
CORAVEL}\ non--variability, spectral type from F8 to M0. 
Both lists are now forming the main programme with 330 objects. 

\begin{figure}[t!]    
\psfig{figure=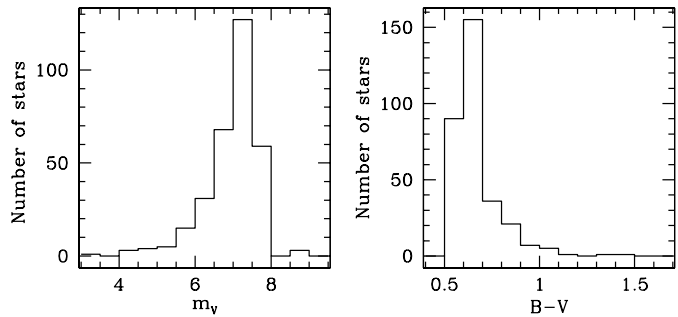,width=\hsize}
\caption{
\label{diagrammVBV}    
The histograms of  $m_{\rm V}$\ and $B-V$\ for the {\footnotesize ELODIE} main 
programme sample stars
}
\end{figure} 

In addition to these main programme stars, for follow--up or phase--coverage reasons 
based on the large access one can have to a 2\,m--class telescope observing time, the 
sample was more recently complemented with 42 stars common with other programmes, 
among which 32 stars, with V magnitudes up to 9, were identified as variable by the 
Keck {\em G--Dwarf Planet Search} \citep{Latham00}.
One star, {\footnotesize HD}\,209458, was in our main programme and now also belongs 
to the collaborative effort with the {\em G--Dwarf Planet Search}.

The {\em Geneva Southern Planet Search programme}  and the {\it {\footnotesize 
ELODIE} Planet Search Survey} have distinct stars samples, based on the declination, 
although the former may occasionally support the monitoring of a usually faint, low 
declination {\footnotesize ELODIE} programme star. 

The planetary companion detection is based on the stellar radial--velocity reflex 
motion produced by gravitational interaction between companion and star. 
The {\footnotesize ELODIE} spectrograph, operating between 3850 and 6800\,\AA, has a 
spectral resolution of about 42\,000. The planet survey makes use of its dual--fiber 
mode where a star and a reference thorium--argon calibration lamp are simultaneously 
observed to monitor and correct the instrumental drift and to achieve the required 
high precision. 
The signal--to--noise ratio per pixel for $m_{\rm V}$ = 7.65 stars is about 100 for 
a 15-minute observation under good atmospheric conditions at {\footnotesize OHP}.
In order to achieve the most homogeneous observations, some exposures are duplicated 
in order to reach the target signal--to--noise. None, except for some special cases, 
are longer than 15 minutes in order that no differential instrumental drift between 
the two beams dominate the resultant accuracy. The total exposure time per source is 
dictated by the goal that the photon--noise be lower than the instrumental limit. 
The radial--velocity measurement is produced on--line by the {\footnotesize ELODIE} 
automatic reduction software which cross-correlates the observed stellar spectrum with 
a binary mask template \citep{Baranne96}. 
Some improvements have been implemented in 2001 in those regions of the mask 
particularly rich in telluric absorption features (see for details \citealt{Pepe02}) 
leading to a current instrument--limited precision better than {\bf 6}\,m\,s$^{\rm -1}$.
 
\begin{figure}[t!]    
\psfig{figure=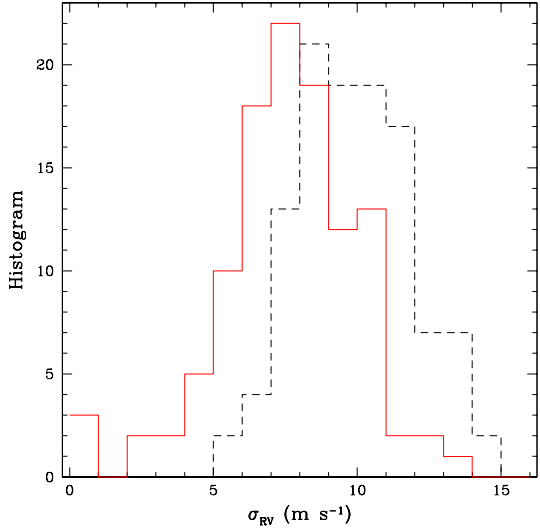,width=0.8\hsize}    
\caption{
\label{hrms}
Histogram of {\footnotesize ELODIE} radial-velocity weighted {\sl rms} for constant stars. The 1723 measurements of 
111 stars having $P(\chi^{\rm 2})$\,$\geq$\,0.05 and $N_{\rm meas}$\,$\geq$\,5 have been used for the plot. These 
selection criteria are sufficient for eliminating highly active stars but lower activity objects can still be 
present in the sample.
The solid curve represents the distribution of the weighted {\sl rms} corrected (quadratically) for the 
photon noise. The main error sources still present in these corrected dispersions are the instrumental error and 
the stellar jitter. The median of this distribution is 7.8\,m\,s$^{\rm -1}$ and 20\% of the stars 
in this sample have weighted {\sl rms} lower than 6\,m\,s$^{\rm -1}$. 
This latter value value is a good estimate of the {\footnotesize ELODIE} instrumental error.
We show for comparison the distribution 
obtained without correcting for the photon noise (dashed line).
}
\end{figure}

\begin{figure}[t!]    
\psfig{figure=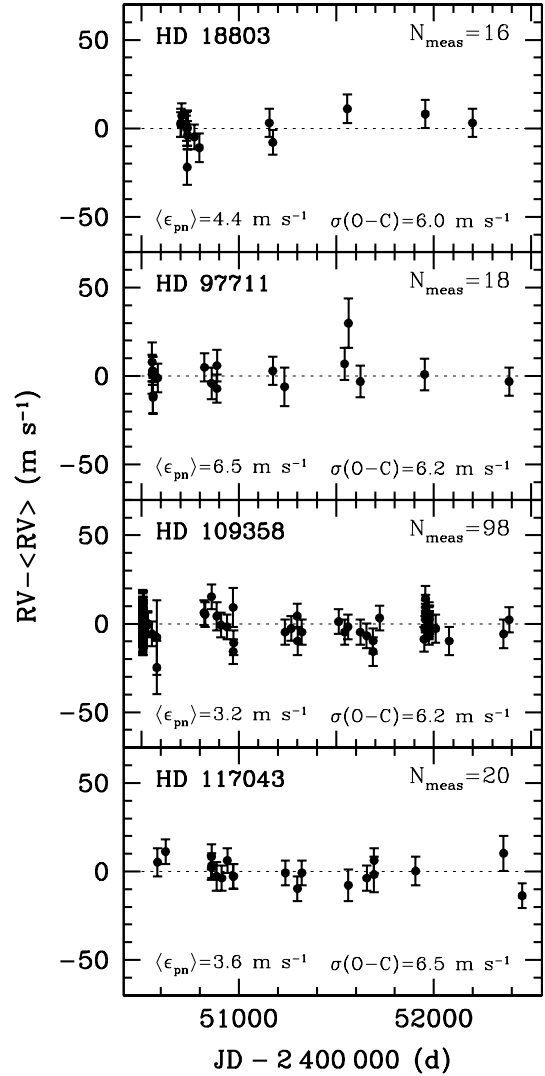,width=0.8\hsize}    
\caption{
\label{precision}
Radial-velocity measurements in km\,s$^{\rm -1}$ versus time for four standard 
stars of the {\footnotesize ELODIE} programme sample: \object{{\footnotesize HD}\,18803}, 
\object{{\footnotesize HD}\,97711}, \object{{\footnotesize HD}\,109358} and \object{{\footnotesize HD}\,117043}.
The mean photon noise $\langle\epsilon_{pn}\rangle$ is indicated at the lower left of each frame and the 
measurements photon 
noise-weighted rms error at the lower right. 
The time span of these measurements is similar to that for the six stars with 
planet candidates.
}
\end{figure}

The improved precision is demonstrated by the distribution of the weighted 
radial-velocity {\sl rms} measured for 111 constant stars in our sample (Fig.~\ref{hrms}).
We corrected the measured dispersions for the contribution of the photon noise. 
The instrumental error can be well estimated by the {\sl rms} values of the low-dispersion tail of 
the histogram.
The median of this distribution is 7.8\,m\,s$^{\rm -1}$ and we find that 20\% of these 
constant stars exhibit a velocity dispersion lower than 6\,m\,s$^{\rm -1}$. 
We also show in Fig.~\ref{precision} our data for 4 constant stars with dispersions 
lower than 6.5\,m\,s$^{\rm -1}$ (photon noise included). Finally, 
the residuals to the orbital solution obtained with {\footnotesize ELODIE} for \object{70\,Vir} 
\citep{NaefeloIII} are 6.1\,m\,s$^{\rm -1}$ (photon noise included). The photon noise 
error of the \object{70\,Vir} measurements is $\epsilon_{\rm phot}$\,=\,2.8\,m\,s$^{\rm -1}$. 
The instrumental error inferred from this long series of values is: 
$\epsilon_{\rm instr}$\,=\,(6.1$^{\rm 2}$\,$-$\,2.8$^{\rm 2}$)$^{\rm 1/2}$\,=\,5.4\,m\,s$^{\rm -1}$.

More recent improvements in the telluric absorption features treatment have allowed us 
to reduce systematic yearly velocity effects that were still visible for some stars in our sample. 
This is for example the case for \object{{\footnotesize HD}\,33636} where the residuals to the fitted orbit
could be reduced by a factor of almost two.


A total of more than 9000 measurements, spread over observing periods that are 
programmed about once every month, have been performed since the beginning of the 
programme at {\footnotesize OHP}. 
All stars but a few ones have been observed more than 10 times and in average 
almost 20 times but those presenting indications of variations may have been 
observed more than 50 times.
 
Table~\ref{decouvertesElodie} lists all the planet candidates discovered so far 
with {\footnotesize ELODIE} having $m_{\rm 2}\sin i$\,$\leq$\,11\,M$_{\rm Jup}$ 
including the six presented in this paper and four others also recently announced. 
More massive companions, with minimum masses in the range 20--75\,M$_{\rm Jup}$, 
among which \object{{\footnotesize HD}\,127506}, \object{{\footnotesize HD}\,174457} 
and \object{{\footnotesize HD}\,185414}, have been discovered with {\footnotesize 
ELODIE} \citep[][ Naef et al. in prep]{Naefman}. 
A peculiar case, \object{{\footnotesize HD}\,166435}, has also emerged from the 
{\it {\footnotesize ELODIE} Planet Search Survey} observations. 
This star presents a chromospheric activity stable enough to mimic a short period 
Keplerian signature. Therefore, any radial--velocity variation potentially due to a 
short period companion is now first checked against such activity using line bisector 
variations, that have been shown to be a robust diagnostic by \citet{Queloz01}.

The list in Table~\ref{decouvertesElodie} includes 
\object{{\footnotesize HD}\,209458\,b} \citep{Mazeh00}, the only extra--solar planet 
for which transits were observed.  
Photometric transits were obtained \citep{Charbonneau00} with timing predictions 
computed from {\footnotesize ELODIE} monitoring and communicated to D.~Charbonneau 
and collaborators. The first spectro-photometric transit was observed with 
{\footnotesize ELODIE} \citep{Queloz00}. 
The photometric and spectroscopic transits yielded notably the planet radius, the 
rotation direction with regard to that of its primary and the mean density 
(0.3\,g\,cm$^{\rm -3}$), essential to assess the gaseous nature of the planets. 
Two planet candidates in Table~\ref{decouvertesElodie} were found around  stars 
not belonging to the  {\footnotesize ELODIE} main programme: 
\object{{\footnotesize HD}\,178911\,Bb}, a planetary companion in a stellar triple 
system \citep{Zucker02}, and \object{{\footnotesize HD}\,80606\,b}, the extra--solar 
planet having the highest orbital eccentricity known so far \citep{Naef01}. 
They are both belonging to the {\em G-Dwarf Planet Search} programme carried out at 
the Keck telescope. Also, \object{{\footnotesize Gl}\,876\,b}, which is presently in 
the closest extra--solar planetary system known, was discovered as a by-product of an 
M--dwarf multiplicity statistical study carried out with the 
{\footnotesize OHP}--{\footnotesize ELODIE} and 
{\footnotesize ESO}--{\footnotesize FEROS} spectrographs and with the 
Canada--France--Hawaii Telescope ({\footnotesize CFHT}) {\footnotesize PUE'O} adaptive optics imaging system by a 
Grenoble--Geneva team \citep{Delfosse98}, and independently by \citet{Marcy98}. 
Later, the presence of two planets on resonant orbits with periods in the ratio 2:1 
that were masqueraded by a unique, more eccentric one, was recognized by 
\citet{Marcy01}.

\begin{figure}[th!]     
\psfig{figure=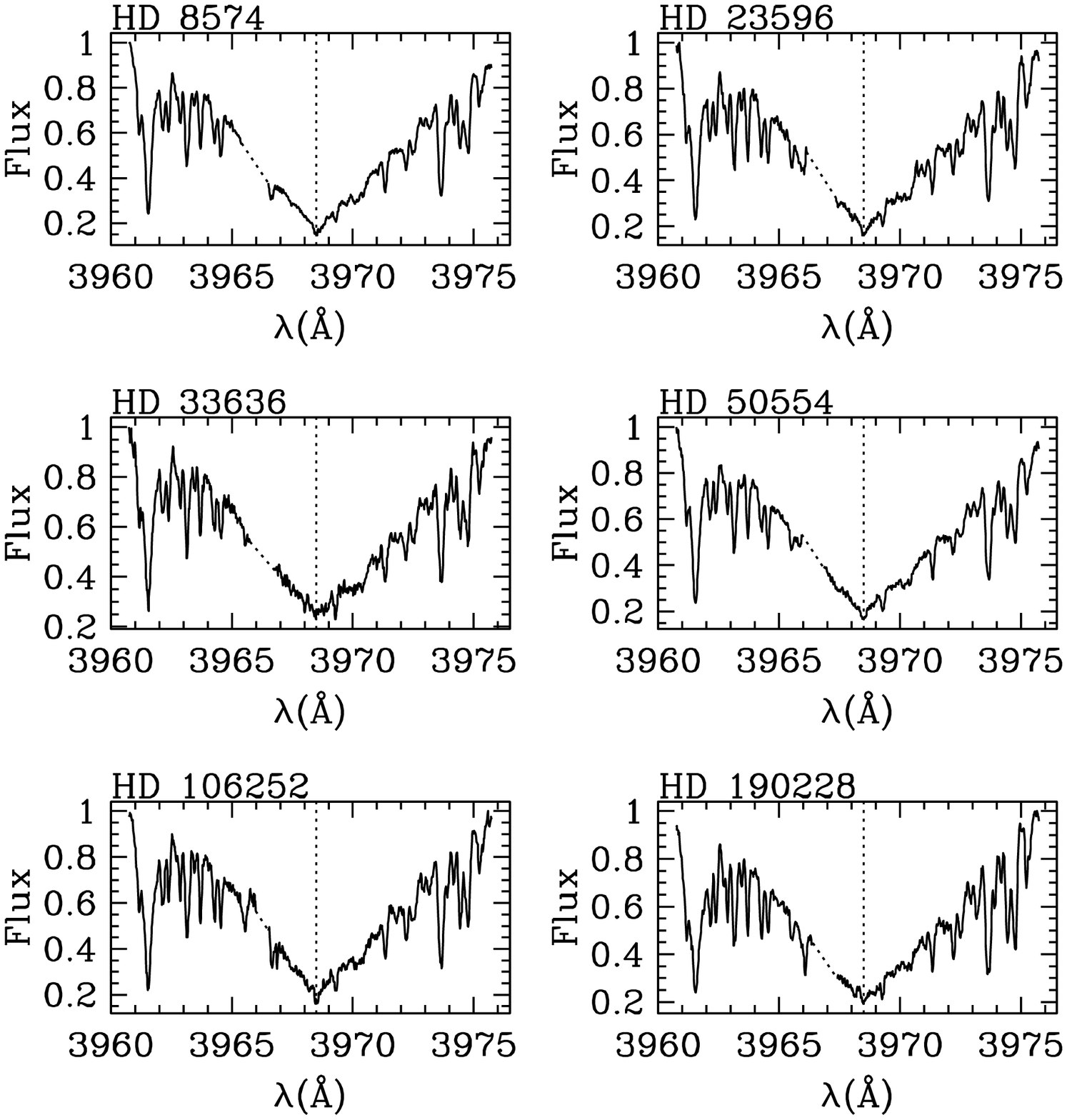,width=\hsize}    
\caption{
\label{raieCalcium}    
The \ion{Ca}{ii}~H line region for the six stars extracted from the 
co-addition of all spectra available for each star. The 3968.49 \AA \ion{Ca}{ii}~H 
line is indicated by a dotted line. None of the stars exhibits any clear trace 
of chromospheric emission feature in the line core (see text).
}
\end{figure} 

\begin{table*}[th!]
\caption{Observed and inferred stellar parameters for the six stars
}
\begin{tabular}{llr@{  $\,\pm\,$  }lr@{  $\,\pm\,$  }lr@{  $\,\pm\,$  }lr@{  $\,\pm\,$  }l}
\hline\hline  \noalign{\vspace{0.025cm}}
\multicolumn{2}{c}{}                              & \multicolumn{2}{c}{\object{{\footnotesize HD}\,8574}}   & \multicolumn{2}{c}{\object{{\footnotesize HD}\,23596}} & \multicolumn{2}{c}{\object{{\footnotesize HD}\,33636}} \\
\noalign{\vspace{0.025cm}}  \hline
{\footnotesize HIP}         &                     & \multicolumn{2}{c}{6643}                                & \multicolumn{2}{c}{17747}                              & \multicolumn{2}{c}{24205}\\
$Sp.\,Type$                 &                     & \multicolumn{2}{c}{F8}                                  & \multicolumn{2}{c}{F8}                                 & \multicolumn{2}{c}{G0}\\
$m_{\rm V}$                 &                     & \multicolumn{2}{c}{7.12}                                & \multicolumn{2}{c}{7.25}                               & \multicolumn{2}{c}{7.00}\\
$B-V$                       &                     & 0.577     & 0.011                                       & 0.634 & 0.009                                          & 0.588     & 0.016\\
$\pi$                       & (mas)               & 22.65     & 0.82                                        & 19.24 & 0.85                                           & 34.85     & 1.33\\
$Distance$                  & (pc)                & 44.2      & $^{1.7}_{1.5}$                              & 52.0  & $^{2.4}_{2.2}$                                 & 28.7      & 1.1\\
$\mu _{\alpha}\cos(\delta)$ & (mas yr$^{\rm -1}$) & 252.59    & 0.76                                        & 53.56 & 0.68                                           & 180.83    & 1.07\\
$\mu _{\delta}$             & (mas yr$^{\rm -1}$) & $-$158.59 & 0.55                                        & 21.06 & 0.55                                           & $-$137.32 & 0.64\\
$M_{\rm V}$                 &                     & \multicolumn{2}{c}{3.89}                                & \multicolumn{2}{c}{3.67}                               & \multicolumn{2}{c}{4.71}\\
$B.C.$                      &                     & \multicolumn{2}{c}{$-$0.034}                            & \multicolumn{2}{c}{$-$0.029}                           & \multicolumn{2}{c}{$-$0.046}\\
$L$                         & (L$_{\sun}$)        & \multicolumn{2}{c}{2.25}                                & \multicolumn{2}{c}{2.76}                               & \multicolumn{2}{c}{1.07}\\
$T_{\rm eff}$               & (K)                 & 6080      & 50                                          & 6125  & 50                                             & 5990      & 50\\
$\log g$                    & (cgs)               & 4.41      & 0.15                                        & 4.29  & 0.15                                           & 4.68      & 0.15\\
$\xi _{\rm t}$              & (km\,s$^{-1}$)      & 1.25      & 0.10                                        & 1.32  & 0.10                                           & 1.22      & 0.10\\
$[$Fe/H$]$                  &                     & 0.05      & 0.07                                        & 0.32  & 0.05                                           & $-$0.05   & 0.07\\
$W_{\lambda, {\rm Li}}$     & (m\AA)		  & \multicolumn{2}{c}{52.1}                                & \multicolumn{2}{c}{76.7}                               & \multicolumn{2}{c}{49.5}\\
$\log n({\rm Li})$          &                     & 2.56      & 0.09                                        & 2.81  & $^{0.11}_{0.07}$                               & 2.46      & $^{0.09}_{0.10}$\\
$v\sin i$                   & (km\,s$^{-1}$)      & 4.04      & 0.61                                        & 3.59  & 0.59         & 2.79                            & 0.65\\
$M_{\ast}$                  & (M$_{\sun}$)        & \multicolumn{2}{c}{1.17}                                & \multicolumn{2}{c}{1.30}                               & \multicolumn{2}{c}{1.12}\\
\noalign{\smallskip}  \hline  \noalign{\vspace{0.025cm}}
\multicolumn{2}{c}{}                              & \multicolumn{2}{c}{\object{{\footnotesize HD}\,50554}} & \multicolumn{2}{c}{\object{{\footnotesize HD}\,106252}} & \multicolumn{2}{c}{\object{{\footnotesize HD}\,190228}}\\
\noalign{\vspace{0.025cm}}  \hline
{\footnotesize HIP}         &                     & \multicolumn{2}{c}{33212}                              & \multicolumn{2}{c}{59610}                               & \multicolumn{2}{c}{98714}\\
$Sp.\,Type$                 &                     & \multicolumn{2}{c}{F8}                                 & \multicolumn{2}{c}{G0}                                  & \multicolumn{2}{c}{G5IV}\\
$m_{\rm V}$                 &                     & \multicolumn{2}{c}{6.84}                               & \multicolumn{2}{c}{7.41}                                & \multicolumn{2}{c}{7.30}\\
$B-V$                       &                     & 0.582    & 0.008                                       & 0.635     & 0.007                                       & 0.793    & 0.006\\
$\pi$                       & (mas)               & 32.23    & 1.01                                        & 26.71     & 0.91                                        & 16.10    & 0.81\\
$Distance$                  & (pc)                & 31.0     & $^{1.0}_{0.9}$                              & 37.4      & $^{1.3}_{1.2}$                              & 62.1     & $^{3.3}_{3.0}$\\
$\mu _{\alpha}\cos(\delta)$ & (mas yr$^{\rm -1}$) & $-$37.29 & 0.93                                        & 23.77     & 0.91                                        & 104.91   & 0.37\\
$\mu _{\delta}$             & (mas yr$^{\rm -1}$) & $-$96.36 & 0.50                                        & $-$279.41 & 0.53                                        & $-$69.85 & 0.57\\
$M_{\rm V}$                 &                     & \multicolumn{2}{c}{4.38}                               & \multicolumn{2}{c}{4.54}                                & \multicolumn{2}{c}{3.33}\\
$B.C.$                      & 		          & \multicolumn{2}{c}{$-$0.038}                           & \multicolumn{2}{c}{$-$0.061}                            & \multicolumn{2}{c}{$-$0.176}\\
$L$                         & (L$_{\sun}$)        & \multicolumn{2}{c}{1.45}                               & \multicolumn{2}{c}{1.27}                                & \multicolumn{2}{c}{4.31}\\
$T_{\rm eff}$               & (K)                 & 6050     & 50                                          & 5890      & 50                                          & 5360     & 40\\
$\log g$                    & (cgs)               & 4.59     & 0.15                                        & 4.40      & 0.15                                        & 4.02     & 0.10\\
$\xi _{\rm t}$              & (km\,s$^{-1}$)      & 1.19     & 0.10                                        & 1.06      & 0.1                                         & 1.12     & 0.08\\
$[$Fe/H$]$                  &                     & 0.02     & 0.06                                        & $-$0.01   & 0.07                                        & $-$0.24  & 0.06\\
$W_{\lambda, {\rm Li}}$     & (m\AA)              & \multicolumn{2}{c}{46.6}                               & \multicolumn{2}{c}{9.7}                                 & \multicolumn{2}{c}{8.8}\\
$\log n({\rm Li})$          &                     & 2.48     & 0.10                                        & 1.62      & $^{0.23}_{0.37}$                            & 1.05     & $^{0.25}_{0.42}$\\
$v\sin i$                   & (km\,s$^{-1}$)      & 3.32     & 0.59                                        & 1.74      & 0.25                                        & \multicolumn{2}{c}{$<$1}\\
$M_{\ast}$                  & (M$_{\sun}$)        & \multicolumn{2}{c}{1.11}                               & \multicolumn{2}{c}{1.02}                                & \multicolumn{2}{c}{0.83}\\
\noalign{\vspace{0.025cm}}  \hline
\end{tabular}
\label{carEtoiles}
\end{table*} 

Fig.~\ref{diagramHR} shows the Hertzsprung--Russell diagram of the 317 stars in the 
{\it {\footnotesize ELODIE} Planet Search Survey} main programme sample for which an 
{\footnotesize HIPPARCOS} entry exists. 
The six stars harboring planet candidates discovered with {\footnotesize
 ELODIE}, presented in this paper, are shown with distinctive symbols. 
$T_{\rm eff}$ is derived from the {\footnotesize HIPPARCOS} $B-V$ by using the 
calibration of \citet{Flower96}\footnote{Note that the coefficients of the 
calibration printed in \citet{Flower96} are incorrect; we rather used the 
coefficients directly obtained from the author.}, except for the six planet-candidate hosts for which the 
effective temperatures listed in Table~\ref{carEtoiles} have been used, and $L$ from {\footnotesize 
HIPPARCOS} apparent magnitude and parallax and the bolometric correction given by 
Flower's calibration. 

In Fig.~\ref{diagrammVBV}, the histograms of $m_{\rm V}$ and $B-V$  for the same 
stars as in Fig.~\ref{diagramHR}, show a clear cut-off in magnitude dictated by the 
{\footnotesize ELODIE} performances and a range in spectral type with a pronounced 
bias toward G0--G5 dwarfs and a strict cut-off at the F8 spectral type.


\section{Stellar properties}\label{secStellarProperties}

The six stars found to harbour planetary companions are 
\object{{\footnotesize HD}\,8574}, \object{{\footnotesize HD}\,23596}, \object{{\footnotesize HD}\,33636}, 
\object{{\footnotesize HD}\,50554}, \object{{\footnotesize HD}\,106252} and \object{{\footnotesize HD}\,190228}. 
Their basic characteristics are summarized in Table~\ref{carEtoiles}. 

Spectral types, apparent magnitudes, color indexes, parallaxes and proper motions 
were all taken from the {\footnotesize HIPPARCOS} catalogue \citep{ESA97}. 
The bolometric corrections of all stars were derived from $B-V$ by using the 
calibration of \citet{Flower96}. 

The atmospheric parameters $T_{\rm eff}$, $\log g$, $\xi _{\rm t}$ and $[$Fe/H$]$ 
 and the stellar masses $M_{\ast}$ are from \citet{Santosstat}.
The projected stellar rotational velocities, $v \sin i$, were obtained using the 
mean {\footnotesize ELODIE} cross-correlation functions ({\footnotesize CCF}) 
dip width \citep{Baranne96}, except for \object{{\footnotesize HD}\,106252}, and the 
calibration by \citet{Queloz98}. 
For \object{{\footnotesize HD}\,106252}, $v \sin i$\ has been derived from 
{\footnotesize CCF}s obtained with {\footnotesize CORALIE} and the calibration in 
\citet{Santos02}. 

Table~\ref{carEtoiles} also gives for each star the equivalent width 
$W_{\lambda,{\rm Li}}$ of the $\lambda 6707.8$\,\AA \ion{Li}{i} line, as measured on 
the co-addition of all the available {\footnotesize ELODIE}\ spectra, corrected for 
the spectrograph straight light.
It can be seen that all stars exhibit a lithium absorption feature which is rather 
strong except for two of them, \object{{\footnotesize HD}\,106252} and 
\object{{\footnotesize HD}\,190228}. 
Lithium abundances, scaled to $\log n({\rm H})$\,=\,12, were derived following the 
method described in \citet{Naefcor5} by using the curves of growth published by 
\citet{Soderblom93} and our tabulated $T_{\rm eff}$.  
The errors were estimated by assuming temperature and equivalent width variations of 
$\pm$\,1$\sigma$.  
The maximum abundance values are obtained by using $T_{\rm eff}$\,+\,1$\sigma$ and 
$W_{\lambda,{\rm Li}}$\,+\,1$\sigma$ and the minimum values correspond to 
$T_{\rm eff}$\,$-$\,1$\sigma$ and $W_{\lambda,{\rm Li}}$\,$-$\,1$\sigma$. 
The resulting errors are not symmetrical because the error on $W_{\lambda,{\rm Li}}$ 
is supposed to be symmetrical and equal to $\pm$\,5\,m\AA.

The \ion{Ca}{ii}~H absorption feature was extracted for the six stars as a 
chromospheric activity diagnostic. 
Fig.~\ref{raieCalcium} displays the spectra of the \ion{Ca}{ii}~H line region 
obtained by the co-addition of all {\footnotesize ELODIE} spectra available for each 
star. 
None of the stars shows an appreciable emission. 
However, in this region, the {\footnotesize ELODIE} spectra do not usually have a 
high signal--to--noise ratio preventing us from the extraction of meaningful values 
of the $\log(R^{'}_{HK})$ indicator. 
The apparent absence of emission does therefore not preclude a mild activity.

\begin{figure*}[th!]    
\psfig{figure=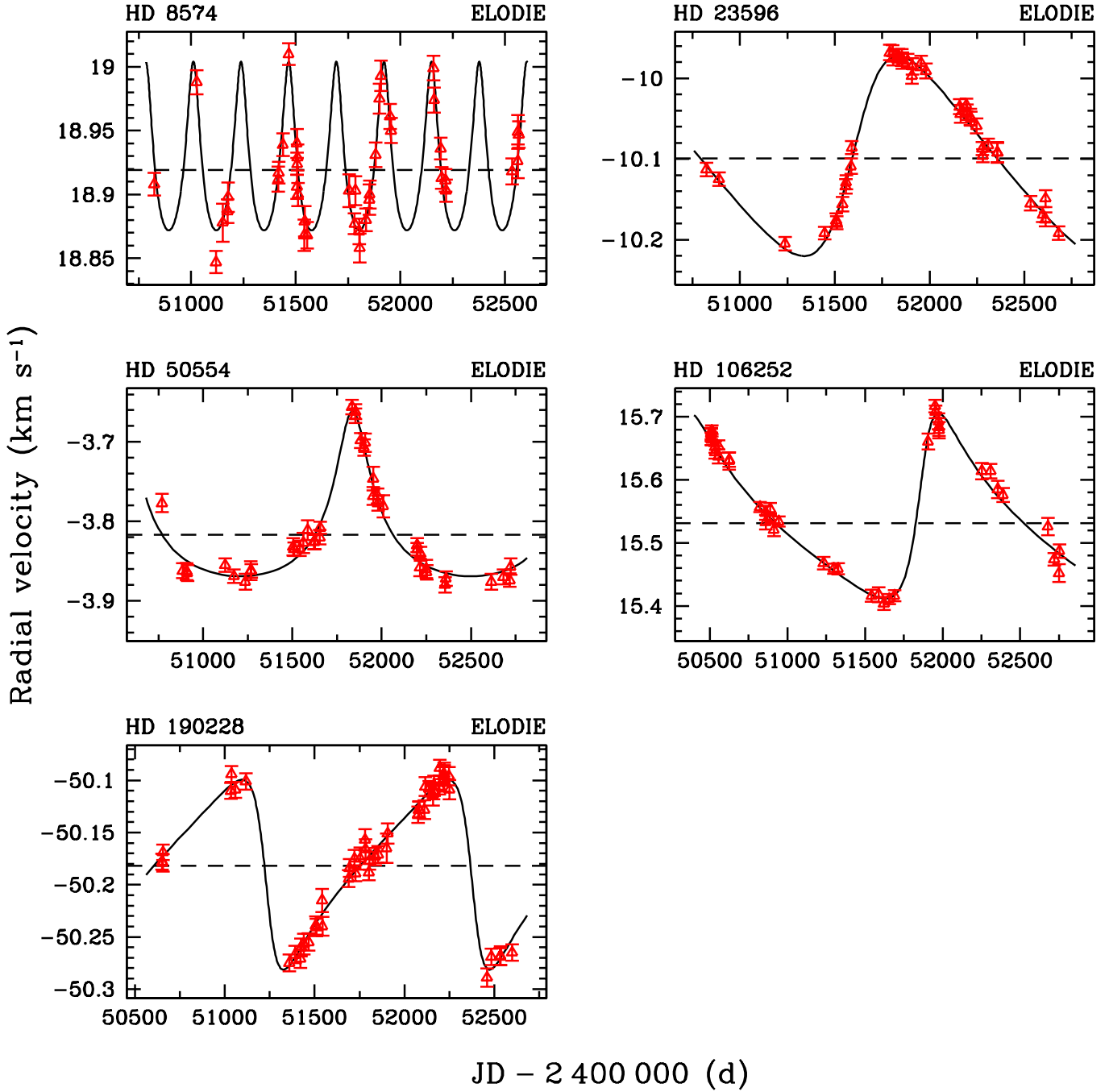,width=1.0\hsize}    
\caption{
\label{vrOrbites_5planets}
{\footnotesize ELODIE} radial--velocity measurements in km\,s$^{\rm -1}$ versus time 
for all stars but \object{{\footnotesize HD}\,33636} shown in Fig.~\ref{vrOrbite_hd33636}. 
Each measurement is plotted with an error bar equal to one estimated standard 
deviation. 
The orbital solution resulting from the best Keplerian fit, listed in 
Table~\ref{solOrbitales} and discussed in the text, is superimposed
}
\end{figure*} 

\begin{figure}[th!]    
\psfig{figure=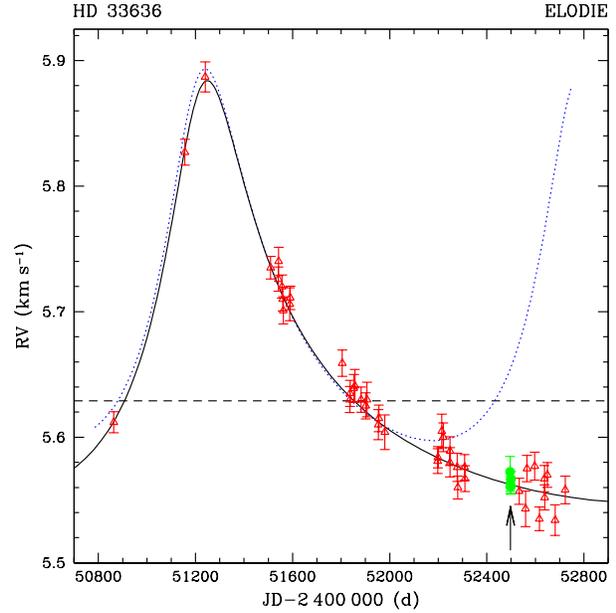,width=0.9\hsize}    
\caption{
\label{vrOrbite_hd33636}
{\footnotesize ELODIE} and {\footnotesize CORALIE} radial--velocity measurements 
in km\,s$^{\rm -1}$ versus time for \object{{\footnotesize HD}\,33636}. 
The 5 {\footnotesize CORALIE} measurements are indicated by the arrow. 
Each measurement is plotted with an error bar equal to one estimated standard 
deviation. 
Since the computed period is not yet covered, the orbital solution is not displayed 
for the whole fitted period. 
The orbital solution resulting from the best Keplerian fit, listed in 
Table~\ref{solOrbitales} and discussed in the text, is superimposed (solid line). 
The orbital solution published by \citet{Vogt02} (dotted line) departs clearly from 
our solution at the epoch of the most recent measurements.
}
\end{figure} 

\section{Orbital solutions}\label{secSolOrbitales}

The achieved precision (photon noise + instrumental error) of the {\footnotesize 
ELODIE} radial--velocity measurements, although improved since the beginning of the 
{\it {\footnotesize ELODIE} Planet Search Survey} (see Sect.~\ref{secElodiePSS}), 
is of the order of 10 m\,s$^{\rm -1}$ for these targets over the time span of the 
reported observations. 
The 254 individual high-precision {\footnotesize ELODIE} radial velocities (+5 {\footnotesize CORALIE} 
measurements) obtained for the six planet-candidate host stars
 are available in electronic form at the {\footnotesize CDS} 
in Strasbourg. 
All these measurements are displayed in Fig.~\ref{vrOrbites_5planets} and 
\ref{vrOrbite_hd33636} which also show the orbital solutions, resulting from the 
best Keplerian fit and described below for each star.

We checked all solutions against false-alarms with Fourier transform analysis 
and found negligible (below the 1\% level) false-alarm probabilities. We also 
checked their fit errors with Monte--Carlo simulations provided within the 
{\footnotesize ORBIT} programme described in \citet{Forveille99}. 
To this end, for each orbital  solution, we ran 10\,000 trials to derive 
Monte--Carlo estimates of the 1 and 3 $\sigma$ confidence intervals. 
None of the solution displays a noticeable discrepancy with these simulations
except in three cases for a few parameters, as mentioned in the text below.

\begin{table*}[t!]
\caption{
\label{solOrbitales}
Fitted orbital elements to the radial--velocity measurements for the six stars, 
characteristics of their planet candidate companions and properties of the fit. 
When necessary, we replaced the fit error by the confidence interval value 
issued from Monte--Carlo simulations (see text).
The $m_{\rm 2}\sin i$ value is based on the $M_{\ast}$ listed in 
Table~\ref{carEtoiles}. 
$N_{\rm meas}$ is the number of measurements that are effectively used in the orbital 
solution fit. 
$\langle\epsilon_{\rm RV}\rangle$ is the mean measurement error and $\sigma({\rm O-C})$ 
is the weighted rms of the residual. 
Comparison of the two values gives an indication of the amount of ""jitter"" which 
can also be appreciated from the $\chi^{2}_{\rm red}$ value 
(the reduced $\chi^{2}$ equal to $\chi^{\rm 2}/\nu$ where $\nu$ 
is the number of degrees of freedom) 
}
\begin{tabular}{llr@{  \,$\pm$\,  }lr@{  \,$\pm$\,  }lr@{  \,$\pm$\,  }lr@{  \,$\pm$\,  }l}
\hline\hline  \noalign{\vspace{0.025cm}}
\multicolumn{2}{c}{}                                                 & \multicolumn{2}{c}{\object{{\footnotesize HD}\,8574}}   & \multicolumn{2}{c}{\object{{\footnotesize HD}\,23596}}  &\multicolumn{2}{c}{\object{{\footnotesize HD}\,33636}$^{\natural}$}  \\
\noalign{\vspace{0.025cm}}
\hline
\noalign{\vspace{0.1cm}}
$P$                               & (days)                           & 227.55        & 0.77                                    & 1565        & 21                                        & 2828         & $^{1090}_{525}$\\
$T$                               & (JD)                             & 2\,451\,467.5 & 6.6                                     & 2\,451\,604 & 15                                        & 2\,451\,211  & 22\\
$e$                               &                                  & 0.288         & 0.053                                   & 0.292       & 0.023                                     & 0.55         & 0.10\\
$\gamma$                          & (km\,s$^{\rm -1}$)               & 18.919        & 0.002                                   & $-$10.099   & 0.002                                     & 5.629        & 0.035\\
$w$                               & ($\degr$)                        & 3.6           & 10.9                                    & 274.1       & 3.9                                       & 340.2        & 6.1\\ 
$K_{\rm 1}$                       & (m\,s$^{\rm -1}$)                & 66            & 5                                       & 124         & 3                                         & 168          & 15\\
$a_{\rm 1} \sin i$                & ($10^{\rm -3}$AU)                & \multicolumn{2}{c}{1.32}                                & \multicolumn{2}{c}{17.1}                                & \multicolumn{2}{c}{36.7}\\
$f_{\rm 1}(m)$                    & ($10^{-9}$${\mathrm M_{\odot}}$) & \multicolumn{2}{c}{5.97}                                & \multicolumn{2}{c}{274}                                 & \multicolumn{2}{c}{822}\\
\noalign{\vspace{0.025cm}} \hline  \noalign{\vspace{0.025cm}}
$m_{\rm 2} \sin i$                & (${\mathrm M_{\rm Jup}}$)        & \multicolumn{2}{c}{2.11}                                & \multicolumn{2}{c}{8.10}                                & \multicolumn{2}{c}{10.58}\\
$a$                               & (AU)        	             & \multicolumn{2}{c}{0.77}                                & \multicolumn{2}{c}{2.88}                                & \multicolumn{2}{c}{4.08}\\
$a_{\rm min}$$^{\dag}$            & (AU)        	             & \multicolumn{2}{c}{0.55}                                & \multicolumn{2}{c}{2.04}                                & \multicolumn{2}{c}{1.83}\\
$a_{\rm max}$$^{\ddag}$           & (AU)        	             & \multicolumn{2}{c}{0.99}                                & \multicolumn{2}{c}{3.73}                                & \multicolumn{2}{c}{6.32}\\
\hline
$N_{\rm meas}$                    &                                  & \multicolumn{2}{c}{41}                                  & \multicolumn{2}{c}{39}                                  & \multicolumn{2}{c}{47}\\
$\langle\epsilon_{\rm RV}\rangle$ & (m\,s$^{\rm -1}$)                & \multicolumn{2}{c}{10.2}                                & \multicolumn{2}{c}{9.2}                                 & \multicolumn{2}{c}{10.4}\\
$\sigma({\rm O-C})$               & (m\,s$^{\rm -1}$)                & \multicolumn{2}{c}{13.1}                                & \multicolumn{2}{c}{9.2}                                 & \multicolumn{2}{c}{9.0}\\
$\chi^{\rm 2}_{\rm red}$          &                                  & \multicolumn{2}{c}{1.99}                                & \multicolumn{2}{c}{1.19}                                & \multicolumn{2}{c}{0.996}\\
\noalign{\vspace{0.025cm}}
\hline  \noalign{\smallskip}	      
\multicolumn{2}{c}{}                                                 & \multicolumn{2}{c}{\object{{\footnotesize HD}\,50554}} & \multicolumn{2}{c}{\object{{\footnotesize HD}\,106252}} & \multicolumn{2}{c}{\object{{\footnotesize HD}\,190228}}\\
\noalign{\vspace{0.025cm}}  
\hline
\noalign{\vspace{0.1cm}}
$P$                               & (days)                           & 1293          & 37                                     & 1600        & 18                                        & 1146        & 16\\
$T$                               & (JD)                             & 2\,451\,832.4 & 15                                     & 2\,451\,871 & 17                                        & 2\,451\,236 & 25\\
$e$                               &                                  & 0.501         & 0.030                                  & 0.471       & 0.028                                     & 0.499       & $^{0.047}_{0.024}$\\
$\gamma$                          & (km\,s$^{\rm -1}$)               & $-$3.817      & 0.002                                  & 15.531      & 0.003                                     & $-$50.182   & 0.004\\
$w$                               & ($\degr$)                        & 355.7         & 4.4                                    & 292.2       & 3.2                                       & 100.7       & $^{2.9}_{3.2}$\\
$K_{\rm 1}$                       & (m\,s$^{\rm -1}$)                & 104           & 5                                      & 147         & 4                                         & 91          & 5\\
$a_{\rm 1} \sin i$                & ($10^{\rm -3}$AU)                & \multicolumn{2}{c}{10.7}                               & \multicolumn{2}{c}{19.1}                                & \multicolumn{2}{c}{8.30}\\
$f_{\rm 1}(m)$                    & ($10^{-9}$${\mathrm M_{\odot}}$) & \multicolumn{2}{c}{96.9}                               & \multicolumn{2}{c}{362}                                 & \multicolumn{2}{c}{58.1}\\
\noalign{\vspace{0.025cm}}
\hline
\noalign{\vspace{0.025cm}}
$m_{\rm 2} \sin i$                & (${\mathrm M_{\rm Jup}}$)        & \multicolumn{2}{c}{5.16}                               & \multicolumn{2}{c}{7.56}                                & \multicolumn{2}{c}{3.58}\\
$a$                               & (AU)		             & \multicolumn{2}{c}{2.41}                               & \multicolumn{2}{c}{2.70}                                & \multicolumn{2}{c}{2.02}\\
$a_{\rm min}$$^{\dag}$            & (AU)		             & \multicolumn{2}{c}{1.20}                               & \multicolumn{2}{c}{1.43}                                & \multicolumn{2}{c}{1.01}\\
$a_{\rm max}$$^{\ddag}$           & (AU)		             & \multicolumn{2}{c}{3.62}                               & \multicolumn{2}{c}{3.97}                                & \multicolumn{2}{c}{3.02}\\
\hline
$N_{\rm meas}$                    &                                  & \multicolumn{2}{c}{41}                                 & \multicolumn{2}{c}{40}                                  & \multicolumn{2}{c}{51}\\
$\langle\epsilon_{\rm RV}\rangle$ & (m\,s$^{\rm -1}$)                & \multicolumn{2}{c}{10.0}                               & \multicolumn{2}{c}{10.6}                                & \multicolumn{2}{c}{8.7}\\
$\sigma({\rm O-C})$               & (m\,s$^{\rm -1}$)                & \multicolumn{2}{c}{11.8}                               & \multicolumn{2}{c}{10.5}                                & \multicolumn{2}{c}{8.0}\\
$\chi^{\rm 2}_{\rm red}$          &                                  & \multicolumn{2}{c}{1.68}                               & \multicolumn{2}{c}{1.21}                                & \multicolumn{2}{c}{0.973}\\
\noalign{\vspace{0.025cm}} \hline
\end{tabular}
\begin{list}{}{}	   
\item$^{\natural}$\ 5 Coralie measurements are included in the solution for HD\,33636 (see text)
\item$^{\dag}$\ at periastron 	   
\item$^{\ddag}$\ at apoastron     
\end{list}
\end{table*}
\subsection{HD\,8574}

We obtained a total of 41 {\footnotesize ELODIE} high-precision radial--velocity 
measurements for \object{{\footnotesize HD}\,8574} which was observed since Jan.~11, 
1998 ({\footnotesize HJD}\,=\,2\,450\,825).  
We list in Table~\ref{solOrbitales} the fitted orbital elements to these measurements 
together with the computed minimum mass $m_{\rm 2}\sin i$\,=\,2.11\,M$_{\rm 
Jup}$ of the planetary companion, assuming a primary star mass of 1.17\,M$_{\sun}$. 
Our discovery of a companion of $m_{\rm 2}\sin i$\,=\,1.95\,M$_{\rm Jup}$ had been 
announced by the April 4$^{th}$ {\footnotesize ESO} Press Release\footnote{{\tt 
www.eso.org/outreach/press-rel/pr-2001/pr-07-01.html}}. 
The orbital solution presented here results from additional measurements obtained 
since then which further improved it. The errors listed in Table~\ref{solOrbitales} are 
derived from Monte--Carlo simulations. They do not differ appreciably from the fit errors assessing 
the quality of this solution despite the rather high value of $\chi^{\rm 2}_{\rm red}$.

The companion of \object{{\footnotesize HD}\,8574} has the lowest period (227.55\,d), well 
constrained (0.3\%) and the best covered in terms of number of orbital cycles of the 
six planet candidates.  
Accordingly, it has also the smallest semi-major axis ($a$\,=\,0.77\,AU, i.e. 17\,mas).  
The slightly abnormal residuals from the fitted orbit yield a moderate "jitter" of 
8.2\,m\,s$^{\rm -1}$, although the primary has no known activity and does not 
exhibit any clear one from the \ion{Ca}{ii}~H line core (Fig.~\ref{raieCalcium}).

\subsection{HD\,23596}

We obtained a total of 39 {\footnotesize ELODIE} high-precision radial--velocity 
measurements for \object{{\footnotesize HD}\,23596} which was observed since Jan.~10, 
1998 ({\footnotesize HJD}\,=\,2\,450\,824). 
Table~\ref{solOrbitales} gives the obtained orbital elements together with the minimum 
mass of the companion ($m_{\rm 2}\sin i$\,=\,8.10\,M$_{\rm Jup}$ assuming 
$M_{\ast}$\,=\,1.30\,M$_{\sun}$).
The computed semi-major axis of the planetary orbit is $a$\,=\,2.88\,AU, i.e. 
55\,mas. 
This solution seems to be well constrained as judged from the absence of abnormal 
residuals and the agreement between fit errors and Monte--Carlo simulations. 

The planet of \object{{\footnotesize HD}\,23596} has the largest semi-major axis  
of the six candidate planets if we put aside the peculiar case of 
\object{{\footnotesize HD}\,33636}. 
With a period of 4.28\,yr, the candidate around \object{{\footnotesize HD}\,23596} belongs to 
the category of long period planets which is naturally growing with the time span 
of the on-going surveys. 
It is also a massive planet. We emphasize the metal-rich nature of the parent star, 
a characteristics not shared by any of the six other stars.

\subsection{HD\,33636}

The fitted orbital elements for \object{{\footnotesize HD}\,33636} to its 42 
{\footnotesize ELODIE} radial--velocity measurements obtained since Feb.~18, 1998 
({\footnotesize HJD}\,= \,2\,450\,863), plus 5 recent {\footnotesize CORALIE} 
measurements, are presented in Table~\ref{solOrbitales}. The velocity offset 
between the two instruments is an additional free parameter. Its fitted value is 
$\Delta RV_{\rm E-C}$\,=\,0.038\,$\pm$\,0.004\,km\,s$^{\rm -1}$.
The variability of the \object{{\footnotesize HD}\,33636} radial velocities has been evident 
for several years since the first measurements were done while the planet was at
its periastron. 
With a primary mass of 1.12\,M$_{\sun}$ given by \citet{Santosstat}, we compute a 
minimum mass of $m_{\rm 2}\sin i$\,=\,10.58\,M$_{\rm Jup}$ and a semi-major 
axis of $a$\,=\,4.08\,AU, i.e.\,142\,mas, for the planet. 
It must be emphasized that the period is certainly not yet covered  
(see Fig.~\ref{vrOrbite_hd33636}), a fact corroborated by the change of the fitted period 
with the addition of the measurements of our last observing campaign. 
At this stage, most of the constraints come from the impact of the eccentricity
 on the radial velocity curve shape. 
Therefore the period of 7.74\,yr is presently badly constrained 
but other parameters should be less affected. 
Interestingly, Monte--Carlo simulations show that the fit errors (listed in 
Table 3) are somewhat overestimated. We adopt the asymmetric Monte-Carlo errors for the period and keep the 
fit errors for the other parameters.

According to our preliminary orbital solution, \object{{\footnotesize HD}\,33636\,b} has the largest 
semi-major axis (4.08\,AU) of the six planets. 
Despite its uncertainty, the period is long and therefore this new planet 
also belongs to the growing category of long period planets. 
It is also be the most massive of our six candidates.

\citet{Vogt02} have announced the independent detection of a companion of
$m_{\rm 2}\sin i$\,=\,7.7\,M$_{\rm Jup}$. 
We have superimposed their orbital solution to ours in 
Fig.~\ref{vrOrbite_hd33636}.
The discrepancy between the two solutions is clearly due to their largely 
underestimated period. Our last campaign provides several 
measurements of good quality with which their orbital curve departs by 6 to 8 
standard deviations of our most recent measurements. Moreover, the period in 
\citet{Vogt02} is shorter than the time span of our measurements.

\subsection{HD\,50554}

A total of 41 {\footnotesize ELODIE} precise radial--velocity measurements were 
obtained for \object{{\footnotesize HD}\,50554} since Nov.~16, 1997 
({\footnotesize HJD}\,=\,2\,450\,769). We list in Table~\ref{solOrbitales} the fitted 
orbital elements together with the computed minimum mass $m_{\rm 2}\sin 
i$\,=\,5.16\,M$_{\rm Jup}$, assuming a primary star mass of 1.11\,M$_{\sun}$. 
The confidence interval resulting from Monte--Carlo simulations is a bit 
larger than the fit error for the period. This is probably due to the peculiar weight 
of the first season measurement in the period estimation (see 
Fig.~\ref{vrOrbites_5planets}). 
We therefore took the confidence interval for the period instead of its fit error.
The semi-major axis of 2.41\,AU, i.e. 77\,mas, is derived for this new planetary 
companion. 
The mean radial--velocity measurement error $\langle\epsilon_{\rm RV}\rangle$ and 
the weighted rms radial--velocity deviation $\sigma({\rm O-C})$ differ notably, 
yielding an appreciable "jitter" of 6.3\,m\,s$^{\rm -1}$. 

Our discovery of a companion of $m_{\rm 2}\sin i$\,=\,4.9\,M$_{\rm Jup}$\ had been 
announced by the {\footnotesize ESO} Press Release nb 07/01 (April 4, 2001). 
The orbital solution presented here results from additional measurements obtained 
since then which further improved it.  

\citet{Fischer02} have confirmed the detection of a companion with $m_{\rm 2}\sin 
i$\,=\,3.7\,M$_{\rm Jup}$. 
Except for their $K_{\rm 1}$ and $\omega$ values that differ significantly
 from ours, the two orbital solutions are compatible. 
The better coverage of the radial--velocity maximum by our measurements probably 
explains the difference. The $m_{\rm 2}\sin i$ discrepancy results from 
the rather large $K_{\rm 1}$ difference between the two solutions.

The \citet{Fischer02} value of the chromospheric activity tracer $\log(R^{'}_{HK})$ ($-$4.94), 
indicating a low activity for \object{{\footnotesize HD}\,50554}, is not useful for 
explaining the observed "jitter", specially since the level of "jitter" is not well 
correlated to the $\log(R^{'}_{HK})$ activity indicator for low-activity stars 
\citep{Saar,Santos00}.

\subsection{HD\,106252}

We gathered a total of 40 {\footnotesize ELODIE} high-precision radial--velocity 
measurements for \object{{\footnotesize HD}\,106252} which was observed since March 
1, 1997 ({\footnotesize HJD}\,=\,2\,450\,509). 
In Table~\ref{solOrbitales} are presented the fitted orbital elements to our 
measurements. 
With a primary mass of $M_{\ast}$\,=1.02\,M$_{\sun}$, we compute a minimum mass 
of $m_{\rm 2}\sin i$\,=\,7.56 M$_{\rm Jup}$ and a semi-major axis of 2.70\,AU, 
i.e. 72\,mas, for \object{{\footnotesize HD}\,106252\,b}. 
This solution is well constrained as confirmed by the absence of abnormal 
residuals. The Monte-Carlo confidence intervals are in good agreement with the fit 
errors.

The discovery of a companion with $m_{\rm 2}\sin i$\,=\,6.8\,M$_{\rm Jup}$\ had 
also been announced by the {\footnotesize ESO} Press Release nb 07/01 (April 4, 
2001). 
The orbital solution presented here results from additional measurements obtained 
since then which further improved it. 
With a period of 4.38\,yr, this new planet also belongs to the growing category of 
long period planets. It is also a rather massive candidate. 

\citet{Fischer02} have confirmed the detection of a companion with $m_{\rm 2}\sin 
i$\,=\,6.96\,M$_{\rm Jup}$, a value in rather good agreement with ours. 
Their period is at more than 3$\sigma$\ from our value.
This difference probably results in their poorer temporal coverage.

The \citet{Fischer02} value of the chromospheric tracer $\log(R^{'}_{HK})$ ($-$4.97), indicating 
a low activity for \object{{\footnotesize HD}\,106252}, agrees with the absence of 
abnormal residuals.

\subsection{HD\,190228}

We obtained a total of 51 {\footnotesize ELODIE} high-precision 
radial--velocity measurements for \object{{\footnotesize HD}\,190228} which was 
observed since Jul.~22, 1997 ({\footnotesize HJD}\,=\,2\,450\,652). 
We list in Table~\ref{solOrbitales} the fitted orbital elements to these 
measurements together with the computed minimum mass $m_{\rm 2}\sin 
i$\,=\,3.58\,M$_{\rm Jup}$\, assuming the primary star mass of 
0.83\,M$_{\sun}$ given by \citet{Santosstat}. 
The semi-major axis $a$\,=\,2.02\,AU, i.e.\,33\,mas, is given as well. 
The confidence interval resulting from Monte--Carlo simulations is somewhat 
larger (indeed asymmetrically) than the fit error for the eccentricity and 
$K_{\rm 1}$. This could be caused by the bad coverage of the periastron (see Fig.~\ref{vrOrbites_5planets}). 
We therefore took the confidence interval for those two parameters instead of 
their fit error.

The discovery of a companion with $m_{\rm 2}\sin i$\,=\,3.4\,M$_{\rm Jup}$\ had 
been announced in \citet{Sivan00}. 
The orbital solution presented here results from additional measurements obtained 
since then which slightly improved it.  

Among the previous discoveries, more than a dozen planet candidates orbit sub-giant 
stars i.e. almost 20\% of the total. 
Not yet statistically meaningful, this ratio however appears to be much larger than 
the proportion of sub-giant stars in the survey sample (19 stars among the {\it 
{\footnotesize ELODIE} Planet Search Survey} sample i.e. 5\% of the total). 
The planet of \object{{\footnotesize HD}\,190228} is therefore not peculiar in this respect. 
On the contrary, \object{{\footnotesize HD}\,190228} is peculiar in terms of metallicity 
since it is the unique case in this paper sample with a metallicity value 
($-$0.24\,$\pm$\,0.06) quite below the solar abundance. 
It is also peculiar with respect to its spectral sub-type, since it is the coolest 
star of this sample. 

\begin{figure}[t!]    
\psfig{figure=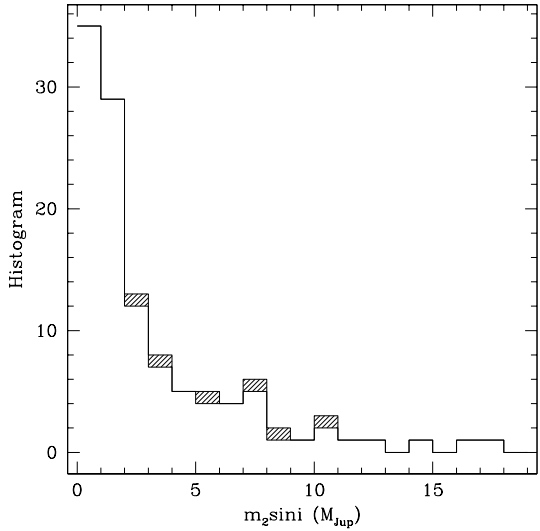,width=0.8\hsize}    
\caption{
\label{histm2sini}
$m_{\rm 2}\sin i$ distribution for the 115 planet candidates discovered so 
far with $m_{\rm 2}\sin i$\,$<$\,18\,M$_{\rm Jup}$. 
The grey square areas point to the contribution of the additional six new 
candidates presented in this paper
}
\end{figure} 


\section{Discussion}\label{discussion}

\subsection{Existence of the companions}\label{existence}

The orbital periods derived for the six planetary companions, spanning from 
0.62 to 4.38\,yr for five of them and even more for \object{{\footnotesize HD}\,33636} 
whose period is fairly imprecise, are much longer than the possible rotation periods 
of the primary stars, even though we have no individual estimation for most of them. 
This makes it very unlikely that any of the radial--velocity curves might be produced 
by activity spot phenomena as it has been observed for the short period variable 
\object{{\footnotesize HD}\,166435} \citep{Queloz01}.  
Nevertheless we checked this for each star by computing the line bisectors of 
the observed {\footnotesize CCF}s and the linear correlation between the 
radial--velocity measurements and the bisector span.
With maximum absolute values of the correlation coefficient below 0.23, we found no 
significant correlation, down to the 1$\sigma$ level, for any of the six stars. 
The interpretation of the Doppler signature by the presence of a companion thus casts
no doubt for the six planet candidates, even for the mildly active stars, 
a conclusion in agreement with that of \citet{Naef00} about long-period planet 
detectability around active stars.

\subsection{$m_{\rm 2} \sin i$\ distribution and mass function}\label{mdistribution}

With six new planet candidates, we can usefully update the distribution of 
$m_{\rm 2}\sin i$. 
Fig.~\ref{histm2sini} displays the histogram of $m_{\rm 2}\sin i$ of the 115 
objects known so far, with $m_{\rm 2}\sin i$\,$<$\,18\,M$_{\rm Jup}$, including our 
six new candidates and about thirty other new candidates announced very 
recently. 
These additional objects, half of which below $m_{\rm 2}\sin i$\,=\,2\,M$_{\rm Jup}$\, 
globally tend to smooth the cumulative frequency distribution and to fill some gaps 
present about $m_{\rm 2}\sin i$\,=\,8\,M$_{\rm Jup}$ (see Fig.~1 of 
\citet{Jorissen01}). 
Because the techniques used to estimate the deconvolved $M_{\rm 2}$ distribution 
(e.g. \citealt{Jorissen01}) is sensitive to the sampling quality, one can infer 
that such an approach will be even more robust with the new, larger input frequency 
distribution. 
More specifically, the latter is likely to result in a smoother $M_{\rm 2}$ 
distribution, specially around and above 10\,M$_{\rm Jup}$ that could make it less 
obvious that 10\,M$_{\rm Jup}$ would be a quasi upper $M_{\rm 2}$ limit. 
Since this is only qualitative, it will be very interesting to know whether this 
study, applied to the new extra--solar planet sample, would rather lead to a 
vanishing distribution in the 15\,M$_{\rm Jup}$ region.

\begin{figure}[t!]    
\psfig{figure=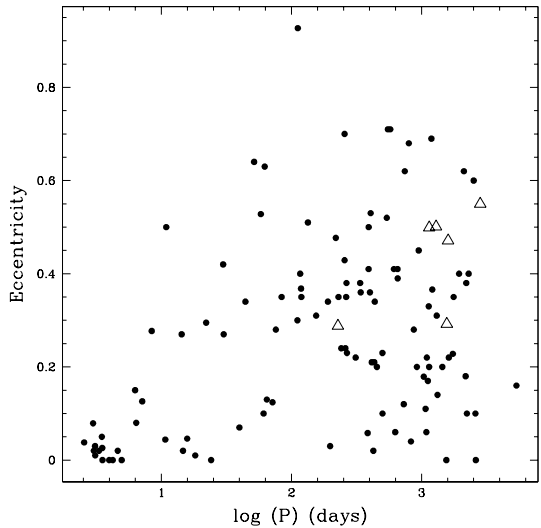,width=0.8\hsize}    
\caption{
\label{diagramEP}
The location of the 115 planet candidates discovered so far with $m_{\rm 2} 
\sin i$ $<18$ M$_{\rm Jup}$ in the diagram eccentricity versus period. 
The six new candidates are shown as open triangles
}
\end{figure} 

\subsection{Eccentricity versus period diagram}

As seen above, the periods of all the companions are long, exceeding three years except 
for \object{{\footnotesize HD}\,8574}. 
Their orbital eccentricities are close to 0.5 for four of the objects and about 
0.3 for the two other ones. 
We have placed these {\bf six} companions on a $e$ vs $P$ diagram together with all 
the other known companions with $m_{\rm 2}\sin i$\,$<$\,18\,M$_{\rm Jup}$ 
(Fig.~\ref{diagramEP}). 
Their location is dictated by the large values of the periods which is the natural 
effect of the bias towards long periods while the time span of the observing 
programmes increases. Apart from this, the diagram shows no particular trend for 
the six companions which are spread in a region of intermediate eccentricities, 
already well populated. 
The locations of our six new companions are in agreement with, and reinforce, 
the general behavior that tends to reproduce the spectroscopic binaries $e$ vs $P$ 
distribution for long periods \citep{MayorIAU200,Mazeh01}, a rather paradoxical 
constraint on planet formation mechanisms.

\subsection{Metallicity}\label{metallicity}

It has been shown that the metallicity distribution of the planet-bearing stars is 
statistically 0.25\,dex higher than for solar neighborhood stars 
\citep[e.g. ][]{Santosmet2}. 
Since four of the six stars with new companions discussed here have solar 
metallicity, they will not help to statistically improve this conclusion. 
Moreover, while \object{{\footnotesize HD}\,23596} is over-metallic by 0.32 dex and therefore 
clearly falls in the upper part of the planet-bearing star distribution of 
\citet{Santosmet2}, \object{{\footnotesize HD}\,190228} is under-metallic by a similar 
amount ($-$0.24\,dex).  
Both stars therefore almost compensates each other's effect on the distribution. 
The six stars will rather serve as additional material for further studies 
of metallicity distribution based on larger statistical samples.

\subsection{Direct imaging feasibility}\label{imaging}

With separations at the orbital elongation larger than 60\,mas, four out of 
the six new planet candidates could in principle be targets for diffraction--limited, 
i.e. adaptive optics, K band near infrared imaging on 8\,m--class telescopes. 
The feasibility of such measurements is of course strongly dependent on the achievable 
contrast and therefore very questionable for long periods, cool Jupiter--like planets. 
Such imaging would anyway place useful constraints on the maximum mass and the orbital 
inclination of the companions. 
Altogether, the objective to image such companions clearly calls for very high 
contrast dedicated instruments \citep[see e.g. ][]{Mouillet01}.


\begin{acknowledgements}    

We acknowledge support from the Swiss National Research Found ({\footnotesize FNRS}), 
the Geneva University and the French {\footnotesize CNRS}. 
We are grateful to the Observatoire de Haute-Provence, the {\footnotesize CFGT} french 
time allocation committee and the {\it Programme National de Plan\'etologie} of 
{\footnotesize CNRS} for the generous time allocation and their continuous support to 
this long--term project. 
We thank T.~Forveille and D.~S\'egransan for the adaptive optics observation and 
data treatment. 
T.~Forveille also helped for the Monte--Carlo simulations with his {\footnotesize 
ORBIT} programme.
We are very much indebted to the Observatoire de Haute-Provence technical and 
telescope support team for the years-long dedication to the {\footnotesize ELODIE} 
operation, specially to A. Vin for his critical software support, and to the various 
night assistants who have taken an essential part in the very demanding telescope 
operation of this observing programme. 
We are grateful to the referee for his comments to the original version of the 
paper. 
This research has made use of the {\footnotesize SIMBAD} database, operated at 
{\footnotesize CDS}, Strasbourg, France.
\end{acknowledgements}

\appendix
\section{HD\,13507}
\begin{figure}[t!]    
\psfig{figure=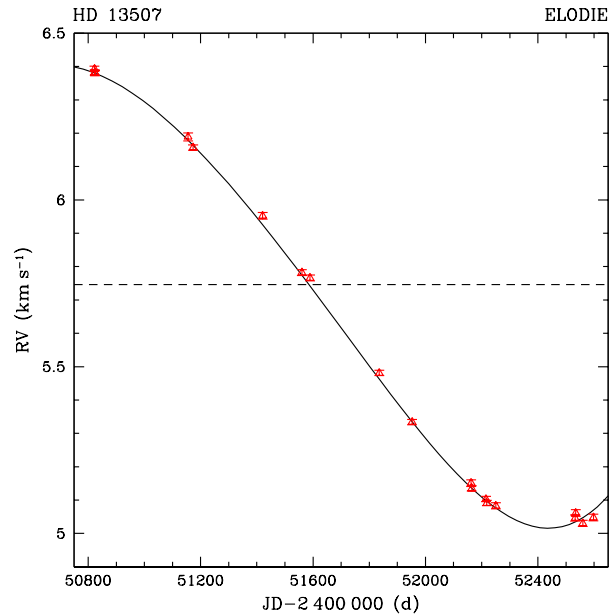,width=0.9\hsize}    
\caption{
\label{orbitHD13507}
The {\footnotesize ELODIE} radial--velocity curve for \object{{\footnotesize HD}\,13507} 
versus time. 
The fit by the reflex motion due to the companion is superimposed to the 
measurements.
}
\end{figure} 

According to the {\footnotesize HIPPARCOS} catalogue, \object{{\footnotesize HD}\,13507} 
is a $m_{\rm V}$\,=\,7.12 G0 dwarf star with a parallax of 38.12\,$\pm$\,0.89\,mas but its absolute magnitude 
$M_{\rm V}$=5.10  $B-V$ (0.672$\,\pm$\,0.007) rather suggest a G4 spectral type. 

The atmospheric parameters and the mass of this star have been derived from a high signal--to--noise 
{\footnotesize ELODIE} spectrum by N.C.~Santos (priv. comm.) 
in the way described in \citet{Santosmet}:
$[$Fe/H$]$\,=\,0.01$\,\pm$\,0.08, $T_{\rm eff}$\,=\,5775$\,\pm$\,80\,K, 
$\log g$\,=\,4.71\,$\pm$\,0.17 (cgs), $\xi _{\rm t}$\,=\,1.33$\,\pm$\,0.11\,km\,s$^{-1}$ and 
$M_{\ast}$\,=\,1.09 M$_{\sun}$.

A total of 19 {\footnotesize ELODIE} precise radial--velocity measurements 
were gathered for \object{{\footnotesize HD}\,13507} since Jan.~8, 1998 
({\footnotesize HJD}\,=\,2\,450\,822). A preliminary but apparently good 
solution, based on a linear drift of 341\,$\pm$\,8\,m\,s$^{\rm -1}$\,yr$^{\rm -1}$ 
and a Keplerian variation compatible with a planetary companion, that was announced 
as such, could be derived until mid 2002.

The later measurements obtained between july and december 2002 (see 
Fig.~\ref{orbitHD13507}) invalidated this interpretation and revealed instead a 
classical spectroscopic binary velocity curve, caused by a low mass star companion. 
This possibly explains the noticed discrepancy between the spectral type from
{\footnotesize HIPPARCOS} and its photometry. 
The computed orbital parameters obtained from a still preliminary fit are: period 
$P$\,$\simeq$\,3000\,days, $e$\,=\,0.14, $K_{\rm 1}$\,=\,0.694\,m\,s$^{\rm -1}$, 
minimum mass of the companion $m_{\rm 2}\sin i$\,=\,52\,M$_{\rm Jup}$, assuming a 
primary star mass of 1.09\,M$_{\sun}$, and semi-major axis $a$\,=\,4.3\,AU, i.e. 
164\,mas. 
This solution is superimposed to the measurements, displayed in 
Fig.~\ref{orbitHD13507}.

We have searched for a companion to \object{{\footnotesize HD}\,13507} by adaptive optics 
imaging. 
An image was obtained in Bracket$\gamma$ with {\footnotesize PUE'O} at the 
{\footnotesize CFHT} on August, $7^{th}$ 2001 ({\footnotesize HJD}\,=\,2\,452\,128.61) 
but no companion shows up at a 3$\sigma$ level. 
A unique negative measurement leaves place to a possible alignment of the components 
but monitoring the binary with high dynamics imaging over a few years should
resolve it and permit to derive the companion mass, a useful information for
studies of sub--stellar objects statistics.

\bibliographystyle{aa}
\bibliography{6planets_V19}
\end{document}